\documentclass[a4paper,12pt]{article}

\usepackage{ifpdf}
\newif\ifpdf
\ifx\pdfoutput\undefined
  \pdffalse
\else
  \pdfoutput=1
  \pdftrue
\fi

\RequirePackage{xspace} %
\RequirePackage{subfigure} %
\RequirePackage[centertags]{amsmath} %
\RequirePackage{amssymb}
\RequirePackage{wrapfig} %
\RequirePackage{calc} %
\RequirePackage{ifthen}
\RequirePackage{tabularx} %
\RequirePackage{flafter} %
\RequirePackage{fancyhdr} %

\ifpdf
  \RequirePackage[pdftex]{color}%
  \RequirePackage{colortbl}%
  \RequirePackage{array}%
  \RequirePackage[pdftex]{graphicx}
  \RequirePackage[ pdftex, plainpages = false, pdfpagelabels,
                 pdfpagelayout = useoutlines,
                 bookmarks,
                 breaklinks = true,
                 linktocpage,
                 pagebackref,                      % to include page numbers in bibliography
                 colorlinks = true,
                 linkcolor = blue,
                 urlcolor  = blue,
                 citecolor = blue,
                 anchorcolor = blue,
                 hyperindex = true,
                 hyperfigures
                 ]{hyperref}

\else
  \RequirePackage{color}
  \RequirePackage{colortbl}
   \RequirePackage{array}
  \RequirePackage[dvips]{graphicx}
  \RequirePackage{hyperref}
  \usepackage{rotating}
\fi

\usepackage{makeidx} %
\usepackage{setspace} %
\usepackage{rotating} %
\usepackage{ecltree}
\usepackage{epic}
\usepackage{supertabular}  %
\usepackage{color}
\usepackage{exscale}
\usepackage{fontenc}
\usepackage{ifthen}
\usepackage{latexsym}
\usepackage{makeidx}
\usepackage{syntonly}
\usepackage{inputenc}
\usepackage{graphicx}
\usepackage{setspace}
\usepackage{caption2}
\usepackage[english]{babel}
\usepackage[square, comma,numbers,sort&compress]{natbib}
\usepackage{hypernat}
\usepackage{boxedminipage}
\usepackage{framed}
\usepackage{longtable}
\usepackage[all]{hypcap}

\newcommand{\note}[1]{\marginpar[left]{\singlespace \tiny #1}}
\renewcommand{\sectionmark}[1]%
      {\markright{\thesection\ #1}} %stops it capitalizing. #1 has value of section name

\renewcommand{\note}[1]{}
\doublespace % \onehalfspace %

\title
{ %
\vspace*{3.0cm} \LARGE{\bf Slip at Fluid-Solid Interface} \vspace*{4.0cm} \\
}

\author{Taha Sochi\footnote{University College London, Department of Physics \& Astronomy, Gower Street, London, WC1E 6BT.
Email: t.sochi@ucl.ac.uk.} \vspace*{5.0cm}}

\begin{document}

\maketitle %
\pagenumbering{arabic}

\newpage
\phantomsection \addcontentsline{toc}{section}{Contents} %
\tableofcontents

\newpage
\phantomsection \addcontentsline{toc}{section}{Abstract} \noindent
{\noindent \LARGE \bf Abstract} \vspace{0.5cm}\\
\noindent %

The `no-slip' is a fundamental assumption and generally-accepted boundary condition in rheology,
tribology and fluid mechanics with strong experimental support. The violations of this condition,
however, are widely recognized in many situations, especially in the flow of non-Newtonian fluids.
Wall slip could lead to large errors and flow instabilities, such as sharkskin formation and spurt
flow, and hence complicates the analysis of fluid systems and introduces serious practical
difficulties. In this article, we discuss slip at fluid-solid interface in an attempt to highlight
the main issues related to this diverse complex phenomenon and its implications.

%%%%%%%%%%%%%%%%%%%%%%%%%%%%%%%%%%%  Head style  %%%%%%%%%%%%%%%%%%%%%%%%%%%%%%%%%%%
\pagestyle{headings} %
\addtolength{\headheight}{+1.6pt}
\lhead[{Chapter \thechapter \thepage}]%
      {{\bfseries\rightmark}}
\rhead[{\bfseries\leftmark}]
     {{\bfseries\thepage}}
\headsep = 1.0cm
%%%%%%%%%%%%%%%%%%%%%%%%%%%%%%%%%%%%%%%%%%%%%%%%%%%%%%%%%%%%%%%%%%%%%%%%%%%%%%%%%%%%

%SSSSSSSSSSSSSSSSSSSSSSSSSSSSSSSSSSSSS
\section{Introduction}

In the flow of viscous fluids over solid surfaces, the `no-slip' at the fluid-solid interface is a
widely-accepted boundary condition with many important consequences. The essence of this condition
is the continuity of the tangential component of the velocity at the wall due to the fact that
viscous fluids stick to the solid surface \cite{BatchelorBook1967, Day1990, Tannerbook2000,
PozrikidisBook2001, FerzigerP2002, KunduC2002}. Strictly speaking, no-slip means that the
instantaneous relative velocity between the fluid and surface, as well as its time average, is zero
\cite{PrieveBook2000}. The absence of slip at the surface creates a boundary layer in the flow
field where the viscous effects and velocity gradients are substantial \cite{Kiljanski1989,
SunE1992, WhiteBook2002, MacielSSGM2002, TiltonBook2007, LamWCJ2007}. This is particularly
important in non-Newtonian fluid systems where shear rate effects, such as shear-thinning, can be
amplified. The thickness of the boundary layer is directly related to the fluid viscosity
\cite{KunduC2002, OertelBook2004}.

Apparently, Stokes was the first to adopt the no-slip as a general condition at the boundary of
solid surfaces \cite{LukMA1987, Day1990}. Although this condition may be described as an assumption
with no hard physical principles \cite{Vinogradova1999, PozrikidisBook2001, TrethewayM2002,
CheikhK2003, ZhangK2004, NetoEBBC2005, SchmatkoHL2005}, several explanations for its physical
foundation have been proposed. These include adsorption of fluid molecules to the solid surface,
and surface roughness at microscopic scale which causes vanishing of wall shear stress
\cite{PozrikidisBook2001, GranickZL2003}. The common factor between most of these proposals is the
presence of forces, which could be described generically as frictional or viscous, that inhibit the
movement of the fluid particles over the solid surface \cite{ChuraevSS1984, PitHL2000, CraigNW2001,
GranickZL2003, OertelBook2004, NetoEBBC2005}. All these explanations can be valid for different
fluid-solid systems and physical conditions; furthermore, more than one of these reasons could
apply in a single flow system simultaneously or under different circumstances. Forces that
influence the adherence of fluids to solid surfaces, and hence determine the effect of stick and
slip, include viscous, elastic, inertial, hydrodynamic, electrostatic, magnetic, gravitational, and
van der Waals \cite{Gennes1985, BonaccursoKB2002, ChoLR2004, HuangGB2006, SethCB2008}. These forces
could play various roles in different flow systems under different circumstances and flow regimes.

There is a general agreement on the restriction of the no-slip condition to viscous flow because
inviscid fluids do not stick to the surface to prevent slip \cite{BatchelorBook1967, Day1990,
FerzigerP2002, KunduC2002}. In fact this justification relies on the assumption that the absence of
viscosity within the fluid itself implies the absence of viscosity relative to the solid surface,
which could be debated because the absence of interaction forces between the fluid molecules does
not necessarily imply the absence of such forces between the fluid molecules and the surface.
Moreover, this justification may not apply if certain nonslip mechanisms, such as disappearance of
wall shear stress due to microscopic roughness, are assumed.

Although the no-slip condition was debated with controversy about its nature and validity in the
nineteenth century, it was eventually accepted as a generally valid condition in rheological and
fluid transport systems with possible exceptions \cite{BatchelorBook1967, CohenM1986,
Schowalter1988, Day1990, ZhuG2001, Denn2001, GranickZL2003, NetoEBBC2005}. Nonetheless, these
exceptions appeared to be so common that its validity became doubtful in many situations. Even when
the no-slip condition seems to hold at macroscopic scale, it could be violated at microscopic and
molecular levels \cite{ThompsonR1990, BizonnJBCRC2002, TrethewayM2002, BonaccursoKB2002,
TrethewayM2003, CurrieBook2003, HervetL2003, ChoLR2004, LichterRM2004, NetoEBBC2005,
SchmatkoHL2005, YanHB2008, Vinogradova1999} with direct impact on microfluidic and nanofluidic
systems \cite{MorrisHG1992, AryaCM2003, LichterMSW2007} and possible ensuing consequences on the
macroscopic behavior \cite{Denn2001}. A prominent exception for the validity of the no-slip
condition is the flow of very low density gas where the mean free path of the molecules are
comparable in size to the dimensions of the flow channel \cite{BatchelorBook1967, Dullien1975,
MorrisHG1992, PrieveBook2000, PozrikidisBook2001, TropeaYF2007}. The effect of the relation between
the mean free path and the dimensions of flow channels on slip can be quantified by the use of
Knudsen number, defined as the ratio of molecular mean free path to a characteristic length scale,
which is mainly used for the flow of gases although it may also be applied to liquids
\cite{KozickiHT1967, Schowalter1988, MorrisHG1992, AryaCM2003, TrethewayM2004}.

The no-slip condition is experimentally verified in numerous cases by direct observation and by
deduction through its inevitable consequences \cite{BatchelorBook1967, LukMA1987,
PozrikidisBook2001, NetoEBBC2005}. Similarly, the slip at wall is confirmed in a large number of
experimental studies (e.g. \cite{ChuraevSS1984, MiglerHL1993, PitHL2000, CraigNW2001, NetoCW2003,
MeekerBC2004a}) for a range of fluids and flow geometries either directly or indirectly through its
implications on the flow such as an unexpected increase in the flow rate or a decrease in the
viscosity or friction factor \cite{LukMA1987}.

The general trend in wall slip when it occurs is that it initiates at a critical shear stress which
characterizes the fluid-solid system under the existing physical conditions \cite{Westover1966,
HatzikiriakosD1991, BerkerV1992, ChenKB1993, Tanner1994, Graham1995, HatzikiriakosHHS1995,
BlackG1996, RosenbaumH1997, PlucinskiGC1998, JoshiLM2000, ChangKS2003, GranickZL2003, GuoWGK2009,
Buscall2010}. It then lasts, or becomes more prominent, over a certain range of deformation rate,
where the magnitude of slip velocity generally varies with the variation of deformation
\cite{LukMA1987, LawalK1998, SanchezVFG2001, Meeten2004}. In some cases, the system response to
reaching the critical shear stress may not be instantaneous, due possibly to the involvement of
viscoelastic or interface effects or to dynamic development of the slip mechanism, and hence the
slip may be delayed by a certain amount of `relaxation' time \cite{HatzikiriakosD1991}. There may
also be points of critical stress where slip increases sharply or transforms from one regime to
another \cite{MiglerHL1993, WangDI1996, LegerHMD1997, Larsonbook1999, DubbeldamM2003, KalyonG2003}.
In shear-thickening fluids, there could be a critical wall shear stress for the slip cutoff
\cite{Kalyon2005}.

In the literature of rheology and fluid mechanics, wall slip is divided into `true slip' where
there is a discontinuity in the velocity field at the fluid-solid interface, and `apparent slip'
where there is an inhomogeneous thin layer of fluid adjacent to the wall with different rheological
properties to the bulk of fluid which facilitates fluid movement \cite{BirdbookAH1987, LukMA1987,
Kiljanski1989, Sorbie1990, KalyonYAY1993, AralK1994, HollandB1995, Barnes1995, GoshawkBKG1998,
ChangKS2003, YeowCKS2004, TropeaYF2007, LamWCJ2007, YanHB2008}. The `apparent slip' designation
stems from the fact that the large velocity gradients across the very thin low-viscosity slip layer
give an impression of slip at wall although the nonslip condition is not violated \cite{CohenM1985,
Barnes1995, Kalyon2005}. Generally, the apparent slip becomes more pronounced as the viscosity of
the slip layer relative to the viscosity of the bulk of fluid decreases \cite{Kalyon2005}.

Although both true and apparent slip have been confirmed experimentally \cite{CohenM1985,
Barnes1995, PitHL1999, HervetL2003, Meeten2004}, in most cases true slip is not the cause of the
tangible macroscopic slip. The reason is that the presence of wall roughness, as a minimum at
microscopic and sub-microscopic levels, and molecular forces between the fluid and solid hinders
substantial movement of the fluid at the interface. Therefore, apparent slip, rather than true
slip, is the more common and viable mechanism for the observable wall slip \cite{SanchezVFG2001}.
Even in the case of yield-stress fluids where slip occurs before bulk-yielding while the material
is still solid, slipping usually occurs through a very thin boundary layer of the fluid where the
stress in this boundary region exceeds the value of yield-stress or through a film from the fluid
phase or from an alien phase. It should be remarked that apparent slip should be extended to
include the case where the slip layer is of a different phase to that of the bulk phase, such as a
thin layer of gas between a liquid and a solid surface or a film of additive substance
\cite{KissiP1996, NetoEBBC2005, TropeaYF2007, Ottinger2008}. The existence of a non-homogeneous
slip boundary layer in complex fluids, such as colloidal dispersions, as required by the definition
of apparent slip has been confirmed by direct experimental evidence \cite{RusselG2000,
MacielSSGM2002, Kalyon2005, LamWCJ2007}.

Although true slip may not be the cause for the tangible slip as observed in the majority of
slippage systems, on microscopic and molecular scale true slip seems inevitable in many
circumstances \cite{PitHL1999, BonaccursoKB2002, CheikhK2003}. Because slippage in these cases has
no measurable impact on the macroscopic flow, it can be ignored in the continuum mechanics
approximations \cite{LiuM1998, GranickZL2003, HenryNEBC2004, LichterRM2004}. Also, the simultaneous
occurrence of true and apparent slip is quite possible where the slip layer itself slips over the
solid surface, possibly with a lower velocity to the apparent slip velocity \cite{LamWCJ2007}, and
hence a hybrid form of true and apparent slip may develop in some systems.

Some statements in the literature may suggest that when slip occurs it is uniform and happens in
all flow zones. Both total slip and nonslip seem to be idealizations \cite{BalasubramanianM1999,
Vinogradova1999} that may occur in special cases and on macroscopic scale. It is more realistic
that what happens in most cases is a local slip which could vary in magnitude from one zone to
another. Local and nonuniform slip, which have some experimental confirmation \cite{RobertDV2004,
GonzalezGSV2009}, can be explained by inhomogeneity of the flow field \cite{BizonnJBCRC2002,
TrethewayM2004}, surface properties and other physical conditions over various flow regions. In
fact, local and nonuniform slip seem inevitable on microscopic level where homogeneity at this
scale can be ruled out because of the involvement of many intricate factors that determine the
situation at this scale. Even on macroscopic scale, wall conditions; such as roughness, coating,
gas pockets \cite{GovardhanSAB2009} and impurities; as well as the fluid properties and ambient
conditions, could vary substantially from one region to another making local and nonuniform slip
unavoidable.

In fact locally varying slip must occur in some circumstances, such as pressure-dependent or
stress-dependent slip, even under totally homogeneous conditions \cite{RuckensteinR1983,
Hatzikiriakos1994, BlackG1996, KumarG1998, TrethewayM2003, HartingKH2006, GuoWGK2009} because of
the presence of pressure or stress gradients as found for example in the capillary flow
\cite{HatzikiriakosHHS1995}. With regards to the flow conditions, locally varying slip seems more
likely to occur in the turbulent systems where the local fluctuations in the flow field are
considerable and may also have time-dependency, making slip itself time-dependent as well as
space-dependent. It should be remarked that spatial dependence of wall slip could complicate the
analysis of flow systems substantially since quantifying the slip contribution will become
difficult or even impossible \cite{BonaccursoBC2003}. Wall slip in such systems would then be
described by averaging the slip-related properties.

Wall slip literature is distinguished by many controversies and contradicting experimental
observations and theoretical conclusions. Slip analysis was pioneered by Mooney \cite{Mooney1931}
who proposed the method of using flow curves from tubes of different sizes to detect and quantify
slip \cite{CohenM1985, CohenM1986, HatzikiriakosD1991, ChangKS2003, ChakrabandhuS2005, LamWCJ2007}.
Wall slip is very complex phenomenon with no well-developed theory to explain it in its variety
\cite{RusselG2000} and provide accurate predictions with solid experimental foundation. One reason
is the difficulty of direct observation and measurement of slip \cite{AryaCM2003, ZhangK2004}.
Another reason is the diversity and complexity of the factors contributing to this phenomenon
especially at microscopic and submicroscopic scales. Another complication of wall slip
investigation is that it could be confused or masked by other overlapping effects; such as shear
banding, polymer degradation, viscous heating, adsorption, entry/exit effects, viscoelasticity and
thixotropy/rheopexy; due to coincidental or causal association between these effects and wall slip
\cite{CohenM1985, BoersmaBLS1991, Graham1995, Barnes1995, Denn2001, BoukanyW2009}. Thus, certain
measures should be taken to avoid confusion with these effects \cite{LamWCJ2007}. Shear banding,
for example, offers an alternative route for stress relaxation and has very similar symptoms to
those of wall slip. Hence, it could compete with wall slip \cite{LettingaM2009} and mask or dilute
its impact making quantitative characterization of wall slip more difficult \cite{Denn2001,
BoukanyW2009}.

Slip at wall is an essential aspect of flow and therefore should not be considered as a casual
anomaly \cite{CohenM1985, Buscall2010}. Wall slip has far-reaching implications on many branches of
science, engineering and industry. These include rheometric measurements, material processing, and
fluid transportation \cite{Westover1966, LukMA1987, Kalyon2005}. Hence, basic understanding of this
phenomenon and developing strategies to deal with it is essential for flow analysis. Despite its
importance, especially for complex fluid systems, slip has been ignored in some studies on systems
which can be strongly influenced by this effect \cite{Buscall2010}. Although this usually
simplifies the experimental procedures and subsequent analysis, it casts a shadow over the validity
of the reported results. Generally, the effect of wall slip in fluid systems is still
underestimated despite the increased awareness of its importance in the recent years
\cite{CohenM1985, Toorman1997, Buscall2010}.

In general, wall slip is an undesirable condition and may be described as nuisance to rheologists
and fluid mechanists on both experimental and theoretical levels \cite{PlucinskiGC1998,
Ottinger2008}. For example, slip is a major source of error in rheometry and rheological
experiments. Due to the fact that common types of commercial rheometers are not designed to account
for wall slip, elaborate and laborious experimental procedures (which include manufacturing one's
own apparatus) are normally required to eliminate or reduce slip. It also introduces serious
complications to the theoretical analysis of fluid systems \cite{Larsonbook1999}. Furthermore, slip
can induce turbulence, flow instabilities and melt fracture, and hence become a source of further
practical and theoretical difficulties \cite{Denn1990, MagninP1990, Graham1995, BlackG1996,
KingL1997, LegerHMD1997, LindnerCB2000, SanchezVFG2001, Denn2001, Georgiou2003, KalyonG2003}.

Nonetheless, wall slip can be advantageous in some situations where its lubricating effect can
promote fluid movement and reduce frictional losses \cite{LukMA1987, HatzikiriakosHHS1995,
Larsonbook1999, ChenDZY2009}. In fact, liquid lubricants are extensively used in various mechanical
systems to enhance solid-solid slip which in most, if not all, cases relies on fluid-solid slip.
Slip can be exploited in some technological and material processing applications to improve surface
finish and material processability \cite{Schowalter1988, HatzikiriakosHHS1995}. Large and
homogeneous slip can also lead to stable flow because it reduces the stress in the sample with an
increase in the flow throughput \cite{Graham1995, BlackG1996, LingGQ2008, LettingaM2009}. The
presence of slip may also improve convective heat transfer at the solid boundary \cite{LukMA1987,
Barnes1995, LawalK1997}. Wall slip has also been proposed as a possible mechanism for drag
reduction \cite{Astarita1965, JonesH1979, MccombBook1992} which has some useful applications,
although other mechanisms seem more viable and may play the major role in this phenomenon with a
possible contribution from wall slip as well. In rheometry, wall slip has been exploited by some
rheologists to measure the yield-stress of viscoplastic materials by observing flow transition on
flow curves \cite{Nguyen1992, Kalyon2005, ChenDZY2009}. In fact such slip-related transition points
on the flow curves, due to change in slip regime and hence slip contribution, can be exploited to
monitor and characterize rheological transition points in general, such as the shift from
shear-thinning to shear-thickening regimes \cite{ChenDZY2009}, due to the variation in the rate of
deformation and ambient conditions.

Slip may be divided into positive, which increases the flow throughput of the flow system, and
negative which decreases the flow throughput. The latter occurs, for example, when the slip layer
is more viscous than the bulk which results in hindering the flow, or when there is a stagnant
layer or a back-flow at wall \cite{KozickiHT1967, MarshallM1967, LiuM1998, PitHL1999}. Positive
wall slip contributes to the total flow throughput, and hence the total flow in slippery systems
has contributions from slip as well as from bulk deformation \cite{BlylerH1970, YilmazerK1992,
KalyonYAY1993, ChakrabandhuS2005, VieraDFG2006}. This may be quantified by the slip fractional flow
rate which is the ratio of the volumetric flow rate of slip contribution to the total flow rate
\cite{CohenM1985, BerkerV1992, YilmazerK1992, ChenKB1993, Kalyon2005, ChenDZY2009}. In some
systems, such as microfluidic and viscoplastic, and over certain deformation rate regimes the
contribution of wall slip to the total flow rate could be substantial \cite{ChuraevSS1984,
CohenM1985, CohenM1986, BerkerV1992, YilmazerK1992, ChenKB1993, AryaCM2003, Kalyon2005,
ChenDZY2009}. In some extreme cases of plug flow, the flow is totally due to wall slip and hence
the fractional flow rate becomes unity.

%SSSSSSSSSSSSSSSSSSSSSSSSSSSSSSSSSSSSS
\section{Slip Factors} \label{Factors}

Direct experimental evidence supported by computer simulations indicate strong correlation between
wall slip and fluid-surface interaction; that is strong interaction prevents or reduces slip
\cite{ThompsonR1990, SunE1992, MiglerHL1993, PitHL2000, BonaccursoKB2002, WallsCSK2003, ZhangK2004,
SchmatkoHL2005, SethCB2008, YanHB2008}. However, it is virtually impossible to quantify this
correlation analytically or empirically or even computationally due to the involvement of many
delicate and interacting aspects, some of which are partially or totally unknown
\cite{ChuraevSS1984, HatzikiriakosD1991, ZhangK2004}. Nonetheless, some general parameters, such as
contact angle and surface tension, have been used as approximate indicators to quantify the
strength of the fluid-solid interaction \cite{SchmatkoHL2005}.

There are several factors that control the fluid-solid interaction and hence determine the onset of
slip and affect its magnitude. One of these factors is the type of fluid and its properties such as
viscosity, yield-stress, elastic modulus, polarity, acidity, electric charge and density
\cite{HuangS1992, AralK1994, HollandB1995, Barnes1995, FrancoGB1998, CraigNW2001, NetoCW2003,
MeekerBC2004a, ChoLR2004, NetoEBBC2005, HartingKH2006, TropeaYF2007}. In the case of
multi-component fluids, such as emulsions, slip can depend on the concentration, particle size
(absolute and relative to the size of flow duct), particle shape, molecular weight, and salinity
\cite{Westover1966, BlylerH1970, KaoNH1975, LukMA1987, HuangS1992, Nguyen1992, MiglerHL1993,
HatzikiriakosHHS1995, FrancoGB1998, Pal2000, SanchezVFG2001, GevgililiK2001, MacielSSGM2002,
WallsCSK2003, MeekerBC2004a, ChakrabandhuS2005, SchmatkoHL2005, TiltonBook2007, LamWCJ2007,
Bertola2007, JossicM2009}. The onset and velocity of slip can also be affected by the presence,
type and quantity of impurities, such as dissolved gases or traces of minerals, in the fluid sample
\cite{KissiP1996, BoehnkeRMWHF1999, GranickZL2003, TrethewayM2003, TropeaYF2007}.

Another factor is the physical and chemical properties of the surface \cite{TrethewayM2004}. Hence,
the material and composition of the surface, as well as the presence of surface active impurities,
can have a strong influence on the onset and magnitude of slip \cite{KraynikS1981, Schowalter1988,
ChenKB1992, ChenKB1993, BlackG1996, LawalK1997, BaudryCTM2001, ChangKS2003, ZhangK2004,
BizonneCSC2005, HartingKH2006, SethCB2008}. The deposition of a very thin film or particles from
another phase, such as a gas or an additive substance, by adsorption or other mechanism can change
the surface properties, such as wettability and roughness, and hence affect wall slip
\cite{RuckensteinR1983, Gennes1985, ChuraevSS1984, KissiP1996, PitHL1999, BoehnkeRMWHF1999,
TyrrellA2001, TrethewayM2003, GranickZL2003, HenryNEBC2004, TrethewayM2004, BizonneCSC2005,
NetoEBBC2005}. The effect of wettability (which is more pertinent to be regarded as a
characteristic of the fluid-solid interaction) and roughness on wall slip will be discussed in the
next two subsections due to their importance and lengthy details. Other surface conditions; like
polarity, electric charge, density, and elasticity; have also been proposed in the literature as
possible reasons for slip enhancement and reduction \cite{ThompsonR1990, Barnes1995,
Vinogradova1999, AryaCM2003, HartingKH2006}.

A third factor that affects slip is the flow regime, as measured by the Reynolds number of the flow
system \cite{LukMA1987}. Slip therefore depends on the rate of deformation and laminar/turbulent
flow conditions \cite{CraigNW2001, ZhuG2001, BonaccursoBC2003, HenryNEBC2004, HartingKH2006,
TiltonBook2007} although some of these dependencies are controversial \cite{CraigNW2001,
NetoCW2003, NetoEBBC2005, HenryNEBC2004}. In most fluid systems, particularly multi-phase
suspensions, slip is more pronounced and dominant at relatively low deformation rates
\cite{CohenM1985, MagninP1987, LukMA1987, MagninP1990, FletcherFLLBe1991, Nguyen1992, Barnes1995,
LiddellB1996, Toorman1997, FrancoGB1998, PlucinskiGC1998, NouarDZ1998, Pal2000, SaakJS2001,
MeekerBC2004a, MeekerBC2004b, Meeten2004, EggerM2006, KelessidisM2008, ChenDZY2009}. This may be
particularly true for shear-thinning fluids where the effect of slip at high deformation rates is
diluted by the drop in the bulk viscosity at these regimes \cite{YilmazerK1992, RusselG2000,
ChenDZY2009}. Also, at high deformation rates the formation of a depleted slip layer in complex
systems can be disrupted because of continuous mixing and possible local turbulent conditions
especially when the surface is rough. In flocculated suspensions, there is also the possibility of
particles breakdown at high rates of deformation with subsequent reduction of slip significance due
to diminished depletion effect \cite{Barnes1995}. In section \S\ \ref{Yield} we will see that the
importance of slip at relatively low deformation regimes is also true in general for yield-stress
fluids \cite{Nguyen1992, MeekerBC2004a, MeekerBC2004b, SethCB2008}, due mainly to the bulk-yielding
on reaching yield-stress and the dominance of bulk flow.

The geometry of the flow system may also induce or reduce slip. For example, the shape and
dimensions of rheometers and the size of the gap width in some rheometric arrangements can have an
impact on the onset and magnitude of slip \cite{GoshawkBKG1998, Pal2000, MeekerBC2004a,
ChangKS2003}. The effect of this factor could be related to its influence on other factors, such as
flow regime and depleted layer thickness \cite{KalyonYAY1993}, or to an increase in the
area-to-volume ratio \cite{ChenKB1993}.

The ambient conditions, such as temperature and pressure \cite{RuckensteinR1983, AralK1994,
Hatzikiriakos1994, HatzikiriakosHHS1995, HollandB1995, BlackG1996, KumarG1998, TrethewayM2003,
HartingKH2006, GuoWGK2009}, also have an impact on the onset and magnitude of slip either through a
change in the properties of fluid and/or surface, or through a change in the nature of fluid-solid
interaction \cite{CohenM1985}. These conditions can also affect another phase which participates in
the slip such as a dissolved species or a thin film of gas or nano-bubbles accreted on the surface
\cite{KissiP1996, TrethewayM2003, TrethewayM2004, TropeaYF2007}. In some cases, slip at wall is
more likely to occur when the fluid is not given sufficient contact ageing time to adhere to the
solid \cite{BlylerH1970, Buscall2010}.

Wall slip may also be influenced by the type of deformation and fluid motion. Oscillatory motion,
for example, could be more susceptible to slip \cite{Pal2000, KhaledV2004} due possibly to an
increased importance of inertial, elastic and time-dependent factors. The interaction between slip
and oscillatory motion, as well as the system response, is generally more pronounced when the fluid
maintains viscoelastic properties \cite{Graham1995}. Also, extensional and shear deformations could
have different impacts on wall slip. The type of driving force, such as being drag or pressure,
although seem to have no effect on slip \cite{Kalyon2005}, it may have an impact on the type of
deformation and hence an indirect impact on the slip. Because some fluids exhibit anisotropic flow
properties, such as certain types of viscoelastic fluids \cite{RosenbaumH1997}, the slip itself
could be anisotropic depending on the direction of the applied pressure or stress.

Specific types of relation between the above-mentioned factors and slip have been reported in the
literature. These include direct relation between slip and viscosity \cite{YilmazerK1992,
BonaccursoBC2003, YanHB2008} or molecular weight \cite{Westover1966, BlylerH1970, Barnes1995,
HatzikiriakosHHS1995, LegerHMD1997, MacielSSGM2002} or temperature \cite{HatzikiriakosD1991,
AralK1994} or pressure \cite{HollandB1995, GuoWGK2009} or particle concentration
\cite{ChakrabandhuS2005, LamWCJ2007} or particle size \cite{Barnes1995}; and reciprocal relation
between slip and pressure \cite{ChenKB1993, HatzikiriakosHHS1995, BlackG1996, KumarG1998,
TrethewayM2003, HartingKH2006, TropeaYF2007} or fluid polarity \cite{ChoLR2004} or density
\cite{HartingKH2006} or viscosity \cite{AralK1994, HollandB1995} or concentration
\cite{SanchezVFG2001} or temperature \cite{ChuraevSS1984}. However, some of these correlations seem
to be system-dependent, as it is obvious from the contradiction in some of these reported
relationships, and hence any generalization requires solid experimental support. Moreover, even if
such specific correlations do exist they may apply under certain circumstances and have a limited
range of validity.

The magnitude and direction of slip is a function of the wall shear stress which is related to the
applied stress \cite{BerkerV1992, ChenKB1993, LawalK1998, MeekerBC2004a}. The onset and velocity of
slip may also depend on the normal component of stress to the wall, as manifested by the observed
dependency of slip on pressure \cite{ChenKB1993, HollandB1995, HatzikiriakosHHS1995, BlackG1996,
KumarG1998, TrethewayM2003, HartingKH2006, TropeaYF2007, GuoWGK2009}. The nature of the
relationship between slip velocity and wall shear stress; such as linear, quadratic, or power law;
is case dependent and is affected by several factors such as the type and state of fluid, flow
regime and physical and chemical properties of the surface \cite{BillerP1987, DubbeldamM2003,
SethCB2008}. However, in many cases a simple direct linear correlation has been assumed where the
proportionality factor is given by a slip coefficient \cite{Mooney1931, ChuraevSS1984,
YilmazerK1992, HollandB1995, Larsonbook1999, GranickZL2003, Kalyon2005, TropeaYF2007}. This simple
linear relationship is essentially the Navier's slip boundary condition \cite{BirdbookAH1987,
Schowalter1988, LawalK1997, Denn2001, TrethewayM2002, MacielSSGM2002, Meeten2004, NetoEBBC2005} for
Newtonian fluids, which is usually expressed as a direct proportionality between the slip velocity
and the shear rate at wall where the proportionality factor is given by a `slip length' or
`extrapolation length' \cite{Gennes1985, MiglerHL1993, Larsonbook1999, Denn2001, MacielSSGM2002,
HervetL2003, AryaCM2003, TrethewayM2004, JosephT2005, LichterMSW2007, GovardhanSAB2009}.

The slip length, which is a simple measure of slip magnitude, is apparently the most commonly used
parameter to characterize slip. It can be interpreted as the extrapolation distance inside the
solid at which the fluid velocity relative to the wall vanishes \cite{MorrisHG1992, MiglerHL1993,
Larsonbook1999, Vinogradova1999, CieplakKB2001, CraigNW2001, ZhuG2001, BonaccursoKB2002,
GranickZL2003, NetoCW2003, GranickZL2003, ChoLR2004, HenryNEBC2004, NetoEBBC2005, SchmatkoHL2005,
VoronovPL2007, YanHB2008}. Therefore the slip length is zero for the nonslip at wall condition.
Although slip length is generally positive, it may also be negative when there is a stagnant layer
at the interface and the nonslip discontinuity lies inside the fluid rather than inside the solid
\cite{Larsonbook1999, HervetL2003}. The existence of such a stagnant layer may be explained by the
presence of a stronger fluid-solid interaction than the interaction within the fluid itself
\cite{HervetL2003}.

The slip length depends on the type of fluid-solid system and its characteristics, such as
viscosity and surface roughness, as well as the ambient physical conditions \cite{Denn2001,
MacielSSGM2002}. Its dependency on the shear rate is controversial with contradicting experimental
results although some of these contradictions can be explained by the proposal that tangible
dependency may be observable only over certain ranges of deformation rate as the slip itself is
normally more pronounced during particular flow regimes \cite{LegerHMD1997, CraigNW2001,
BonaccursoKB2002, HervetL2003, GranickZL2003, TrethewayM2004, SchmatkoHL2005, HuangB2007}.
Moreover, independence of slip length from shear rate, as observed in some experimental studies,
may arise because of the Newtonian nature of the fluid with shear viscosity being independent of
the shear rate, associated with a linearity of relationship between slip velocity and shear stress
\cite{Larsonbook1999}. Anyway, the dependence on shear rate can be system-dependent, and hence the
failure of some investigators to observe this dependency in their systems cannot negate the
confirmation by others in their own systems.

Normally, when the slip length is very small compared to the dimensions of the flow path the effect
of slip diminishes and could become negligible \cite{LegerHMD1997, Larsonbook1999, GranickZL2003,
Meeten2004, GovardhanSAB2009}, although this may not be true in general. The reported values of
slip length, which vary largely depending on the system and the reporter, range between a few
micrometers to a few nanometers and even smaller, although the majority are in the nanometer regime
\cite{Larsonbook1999, CraigNW2001, TrethewayM2002, HervetL2003, GranickZL2003, Meeten2004,
TrethewayM2004, JosephT2005, SchmatkoHL2005, HuangGB2006}. Exceptionally large values in the order
of tens and even hundreds of microns for certain disperse systems can also be found in the
literature \cite{Gennes1985, Larsonbook1999, Denn2001, MacielSSGM2002, GovardhanSAB2009}. These
large slip lengths may be explained by the presence of gaseous films or pockets on the surface
which cause a large apparent slip \cite{TrethewayM2004, NetoEBBC2005, GovardhanSAB2009}. The
existence of such films in some slip systems is supported by direct experimental evidence
\cite{JosephT2005}. When the slip length is large, the slip effect could be significant even when
the flow paths are large \cite{Larsonbook1999}. The slip length in multi-component fluid systems is
generally larger than that in single-component fluid systems \cite{Meeten2004}. Slip length can be
time-dependent \cite{HervetL2003, GovardhanSAB2009} due to time-dependency of the flow field during
transition stages, or because of thixotropic or viscoelastic developments and hence slip length
would depend on the deformation history \cite{HatzikiriakosD1991, Graham1995, FrancoGB1998}.

%SSSSSSSSSSSSSSSSSSSSSSSSSSSSSSSSSSSSS
\subsection{Wettability}

There are contradicting views about the occurrence of slip on totally or partially wetted surfaces
\cite{BizonnJBCRC2002, NetoCW2003, HervetL2003, GranickZL2003, HenryNEBC2004, BizonneCSC2005,
NetoEBBC2005, SchmatkoHL2005, HuangGB2006}. The validity of the nonslip condition when the viscous
fluid does not wet the solid surface is another controversial issue \cite{LukMA1987,
PrieveBook2000, Tannerbook2000}. Apparently, if the `non-wetting' condition implies the absence of
`sticking forces' between the fluid and solid surface at the points of contact under dynamic flow
conditions, possibly due to the presence of stronger counteracting forces from the fluid molecules
\cite{ChuraevSS1984, Vinogradova1999, GranickZL2003, NetoEBBC2005}, then the no-slip condition
should be violated in this case. However, as non-wettability varies in magnitude in the case of
partial wetting, the slip effect should also vary accordingly if such a correlation does exist. In
fact such a correlation has been confirmed in general \cite{LukMA1987, BonaccursoKB2002,
BizonnJBCRC2002, TrethewayM2002, BonaccursoBC2003, BizonneCSC2005, HartingKH2006} although the
details are controversial or obscure \cite{PlucinskiGC1998, HuangGB2006}.

In this context, it has been asserted \cite{ZhangK2004} that experimental and molecular dynamics
simulation studies have demonstrated that slip usually occurs on hydrophobic surfaces. Since
wettability, as inversely measured by the fluid-solid contact angle, indicates the strength of
fluid-solid interaction, slip should be directly related to hydrophobicity \cite{LukMA1987,
NetoEBBC2005}. Although qualitatively this correlation has been validated in a number of studies
\cite{Schnell1956, KaoNH1975, RuckensteinR1983, BaudryCTM2001, GranickZL2003, TrethewayM2004,
NetoEBBC2005, SchmatkoHL2005}, there seems to be no obvious quantitative correlation
\cite{LukMA1987, ChoLR2004, HartingKH2006, VoronovPL2007, TropeaYF2007}.

Anyway, even if hydrophobicity was a sufficient condition for slip it is not a necessary condition,
as wall slip can occur on hydrophilic surfaces as well as hydrophobic surfaces
\cite{BonaccursoKB2002, BonaccursoBC2003, NetoCW2003, HervetL2003, GranickZL2003, HenryNEBC2004,
TrethewayM2004, SchmatkoHL2005, HuangB2007, TropeaYF2007}, and this should be particularly true for
apparent slip. In fact apparent slip, according to some of its mechanisms at least, requires strong
fluid-solid interaction, and hence there should be a reciprocal correlation between non-wettability
and apparent slip. It may be concluded, therefore, that wall slip on non-wetted or partially-wetted
surfaces is basically a true slip. However, some studies indicate that the origin of
hydrophobicity, at least in some cases, is the existence of gas films or nano-bubbles
\cite{RuckensteinR1983, TyrrellA2001, BizonnJBCRC2002, TrethewayM2003, TrethewayM2004,
NetoEBBC2005, Vinogradova1999}, rendering these cases to be an instance of apparent slip. It is
noteworthy that the failure of some investigators (e.g. \cite{TrethewayM2002}) to detect slip in
hydrophilic systems cannot negate the reported confirmations by other investigators.

There are many contradictions in the reported results on the relation between slip and wetting,
which include some extreme findings. For example, some investigators \cite{WallsCSK2003} have
reported experimental results in which hydrophobic surfaces reduced slip significantly while
hydrophilic surfaces had minimal effect on slip, justifying their results by sample-surface
interaction. One reason for these contradictions is that the results come from different systems
where factors other than wettability have contribution to the slip. Since wall slip is very complex
phenomenon in which many delicate factors are involved \cite{GranickZL2003, ZhangK2004},
wettability is not a sufficient condition for nonslip. In fact even if we assume that the
contradicting results come from virtually identical systems and are obtained under very similar
physical conditions, there is still a substantial probability that the systems differ in some
subtle factors, possibly at micro or nano scale, that have a direct impact on the slip.

It should be remarked that non-wettability under static measurement conditions cannot rule out the
possibility of the development of sticking forces under dynamic flow conditions which may prevent
slip. Similarly, dynamic interaction can introduce factors that can encourage slip on totally or
partially wetted surfaces. The absence of obvious correlation between wetting characteristics and
slip in general, as manifested by the contradicting experimental reports, may be partially
explained by this fact because of the possibility of a dynamic interaction in some systems under
flow conditions that alter the relation between wettability and slip. Many studies on the
correlation between wettability and slip try, unjustifiably, to extend the static state interaction
between fluid and surface, as obtained from static wettability measurements, to the dynamic
situation. Other physical factors, such as ageing contact time between fluid and solid, may also
play a role in some situations, hence changing the relation between wettability and slip. As long
as these factors are not absorbed in the definition of wettability and are not considered in the
static measurements used to characterize the dynamic flow situations, the relation between
wettability and slip remains ambiguous.

In conclusion, although wettability generally promotes nonslip condition and vice versa, no
confirmed specific generalization can be made, especially at quantitative level, based on the
reported experimental and theoretical results, because the relation between wall slip and
wettability is very complex matter in which many factors are involved. Notably, the fluid-solid
dynamic interaction could play a significant role in the overall slip behavior and hence the static
characterization may give a partial, and possibly misleading, picture \cite{JosephT2005}. Other
known and unknown factors could also have an impact on the interaction between wettability and wall
slip.

%SSSSSSSSSSSSSSSSSSSSSSSSSSSSSSSSSSSSS
\subsection{Surface Roughness} \label{Roughness}

A prominent example of slip-related surface conditions is wall roughness which normally inhibits
slip \cite{MagninP1990, ChenKB1992, ChenKB1993, Barnes1995, FrancoGB1998, PitHL2000, HervetL2003,
SethCB2008}. However, its effect generally depends on the rate of deformation and the type of
fluid-solid system \cite{MagninP1990, VieraDFG2006}, as well as the nature of roughness itself, and
in some cases it may have limited or nil effect \cite{MagninP1990, ChenKB1992, AralK1994,
GoshawkBKG1998, PlucinskiGC1998, JossicM2009, DivouxTBM2010}. The effect of roughness on slip also
depends on the relative size of the dispersed molecules/particles in the solution compared to the
size of asperities \cite{Nguyen1992, Tanner1994, PlucinskiGC1998, JabbarzadehAT1999, PitHL2000,
MerkakJM2006, DivouxTBM2010} and the dimensions of flow ducts and spacing of rheometric gaps
\cite{Nguyen1992}. In some fluid systems, surface roughness may even enhance wall slip by
encouraging certain slip mechanisms \cite{Nguyen1992, BonaccursoBC2003, NetoEBBC2005,
GovardhanSAB2009}. This enhancement, which may be ascribed to sample fracture \cite{Meeten2004},
could be related to the roughness type. Roughness, or excessive roughness, may also induce
secondary flows, or cause sample fracture which could be particularly critical for viscoplastic
materials below their yield-stress \cite{MagninP1987, AralK1994, ChangKS2003, Meeten2004,
MeekerBC2004a}. In the case of roughness anisotropy, such as parallel grooves in a certain
direction, the slip can be strongly dependent on the direction of flow \cite{Princen1985,
Nguyen1992}.

The effect of roughness on wall slip is a controversial issue with quantitatively and qualitatively
contradicting findings \cite{CraigNW2001, NetoEBBC2005, VieraDFG2006}. Various proposals have been
put forward to explain the effect of roughness on slip in different systems and under various
circumstances. For example, it has been proposed that rough surfaces provide good grip on the
sample hence preventing or reducing slip \cite{Pal2000}. Another proposal is that roughness
disrupts or prevents the formation of a depleted layer \cite{Pal2000, EggerM2006}. In this regard,
it has been argued \cite{TropeaYF2007} that for roughness to be effective by disrupting the
depleted slip layer, the roughness profile must be larger than the thickness of this layer. Surface
roughness can also accommodate pockets or films of another phase (such as a gas phase in a
liquid-solid system) which can alter surface properties, such as wettability, and hence affect slip
\cite{Vinogradova1999, TrethewayM2003, GranickZL2003, NetoEBBC2005, TropeaYF2007,
GovardhanSAB2009}.

Some researchers have attempted to correlate the effect of surface roughness to its impact on
wettability \cite{BizonneCSC2005} by distinguishing between two cases, that is if the fluid totally
wet the surface then roughness reduces slip while if it wets the surface partially then roughness
increases the slip through the formation of trapped pockets or films of a gas or a vapor phase at
the cavities and crevices of the surface. However, the literature of wall slip seems to contradict
the existence of such a simple correlation between roughness and wetting properties of the surface
because roughness, although seems to improve wettability in some systems and reduce it in others,
it does so through a multitude of other obscure factors \cite{NetoEBBC2005}.

The effect of roughness has also been explained by the proposal that on atomic scale it increases
the fluid-solid interaction through an increase in the contact points, while it changes the flow
pattern at larger scales \cite{BonaccursoBC2003}. In this context, it has been stated
\cite{NetoCW2003} that roughness on a molecular scale could suppress slippage, while on a larger
scale it might enhance it. However, this view seems to contradict the conclusions of other
experimental and theoretical studies on this issue. Some authors \cite{Tanner1994} have correlated
the effect of roughness on slip to the fluid molecular size by distinguishing between the case of
small molecules relative to the surface roughness scale where nonslip condition holds, and the case
of large macromolecules where slip could happen. However, these types of correlation seem to depend
on the nature of the slip mechanism, although they may reflect the trend in some categories of
fluid systems.

Roughness is normally quantified by giving the average (or root mean square) height of the
roughness peaks or the depth of grooves \cite{Meeten2004, TokpaviJMJ2009, GovardhanSAB2009,
DivouxTBM2010}, or by giving the peak-to-peak mean distance \cite{BaudryCTM2001, BizonnJBCRC2002,
BizonneCSC2005}. It may also be quantified by other predefined parameters \cite{ChenKB1992,
ChenKB1993, AralK1994} such as the ratio of real to apparent surface area \cite{NetoEBBC2005}.
However, some of these quantifiers may not provide full characterization as they contain an element
of ambiguity, and hence should be regarded as approximate indicators. A number of direct
measurement techniques, such as profilometry \cite{ChenKB1993, AralK1994, GovardhanSAB2009},
scanning electron microscopy \cite{ChuraevSS1984, ChenKB1993, AralK1994}, and atomic force
microscopy \cite{BizonnJBCRC2002, BizonneCSC2005, JosephT2005, SchmatkoHL2005} have been used to
measure and characterize surface roughness.

%SSSSSSSSSSSSSSSSSSSSSSSSSSSSSSSSSSSSS
\section{Slip Mechanisms} \label{Mechanisms}

Slip is a diverse phenomenon with different physical causes that vary depending on the flow system
and surrounding conditions. The fluid-solid interaction is affected by a number of static and
dynamic factors that influence the adherence and slip. There is no general theory to describe wall
slip in all situations; however, there is a number of mechanisms proposed to explain slip in
certain circumstances and for certain flow systems \cite{LukMA1987}. One of these mechanisms is the
depletion of the boundary layer where the particles in a sheared dispersed system, such as
emulsions and suspensions, migrate away from the boundary region resulting in a very thin
low-viscosity layer adjacent to the wall which acts as a lubricating film that facilitates fluid
movement \cite{Vand1948, ChuraevSS1984, CohenM1985, Sorbie1990, BerkerV1992, KalyonYAY1993,
AralK1994, Barnes1995, HollandB1995, LawalK1998, FrancoGB1998, BalmforthbookC2001, SanchezVFG2001,
MacielSSGM2002, VieraDFG2006, ChenDZY2009}. Some statements in the literature may suggest that this
is the only viable mechanism for wall slip. However, it is more appropriate to regard this as the
main slip mechanism for apparent slip in a certain category of fluids, such as suspensions and
emulsions, especially those whose viscosity is highly-dependent on concentration \cite{CohenM1986,
Barnes1995, MeekerBC2004a}. In other systems, apparent slip can arise from other mechanisms;
furthermore, true slip can also occur with no need for a depleted or even a slip layer
\cite{PitHL1999}.

The general assumption in the literature is that the depletion of the boundary layer will lead to a
drop in its viscosity and hence a slip enhancement. However, depending on the nature of correlation
between concentration and viscosity (which in the case of non-Newtonian fluids could also be
affected by shear-induced effects, such as shear-thinning and shear-thickening, in the
highly-deformed boundary region) the depletion could, in theory, lead to an increase, as well as a
decrease, in the viscosity of the boundary layer. Nonetheless, since the viscosity of the
dispersing (continuum) phase is usually very low compared to the viscosity of the disperse system,
the viscosity of the depleted layer will decrease if total or substantial depletion is assumed with
no significant shear-thickening offset, because the viscosity of the slip layer then is the
viscosity of the dispersing phase with negligible impact from the dispersed phase \cite{CohenM1985,
CohenM1986, Kiljanski1989, Sorbie1990, YilmazerK1992, KalyonYAY1993, HollandB1995, Barnes1995,
Pal2000, MacielSSGM2002, GranickZL2003, ChakrabandhuS2005, Kalyon2005, LamWCJ2007, ChenDZY2009}. In
fact the viscosity of the slip layer could even be lower than the viscosity of the dispersing phase
if shear-thinning occurs at the boundary.

There are several explanations for the development of the depleted layer region. For example, in an
elasto-hydrodynamic lubrication model proposed \cite{MeekerBC2004a, MeekerBC2004b} to explain slip
in highly concentrated suspensions of soft particles, the particles-wall interaction during slip
occurs through non-contact elasto-hydrodynamic forces which appear because the relative motion
creates an asymmetric deformation in the elastic particles near the wall resulting in a lifting
force that pushes them away from the wall. Another explanation, which is based on entropic
exclusion, is that the dispersed particles/molecules have more available orientations away from the
surface than near the surface \cite{Sorbie1990, FletcherFLLBe1991, HuangS1992, HollandB1995,
Barnes1995, Kalyon2005} leading to a lower concentration in the boundary region. This has been
criticized by the fact that the slip layer thickness in some systems could be much larger than the
hydrodynamic diameter of the molecules \cite{CohenM1986}. However, this criticism cannot rule out
the validity of this mechanism in other systems where the slip layer thickness is comparable to the
molecular dimensions. Moreover, entropic exclusion in dynamic flow situation could be a
contributing factor to the depletion and hence it does not imply comparability in size. A third
explanation is that the dissolved species diffuses away from the high-stress boundary region to the
low-stress central region leading to the formation of a thin solvent layer \cite{RuckensteinR1983,
LukMA1987, BerkerV1992, JoshiLM2000}. A similar explanation of shear-induced lift, due to inertial
effects or shear gradients, that forces the dispersed particles to migrate away from the solid
surface has also been proposed \cite{AralK1994, Barnes1995, KingL1997, BalmforthbookC2001,
MacielSSGM2002, HuangB2007, TropeaYF2007}.

Various physicochemical forces; such as electrostatic, electrokinetic, steric, osmotic,
hydrodynamic, gravitational, and viscoelastic; can also contribute to the formation of the depleted
layer and its subsequent development \cite{CohenM1985, CohenM1986, Barnes1995}. For example,
gravitational sedimentation or electrodynamic forces between the surface and the particles of the
dispersed phase could reduce the concentration of dispersions near the wall, hence creating a
depleted layer at the boundary or altering its thickness \cite{AralK1994, Barnes1995, Pal2000,
JosephT2005, HuangGB2006}. It should be remarked that depleted layer scenario can be extended to
include single-component fluids where the slip at wall occurs through the development of a sparse
low-viscosity layer at the boundary due to repulsive forces, such as electrostatic, between the
surface and the fluid molecules. This mechanism, which does not require a depletion of a dispersed
phase, is feasible for simple gaseous fluids whose density can experience substantial variations
depending on the physical conditions.

Another proposed mechanism for wall slip, which is also viable for multi-component fluids such as
colloids, is the development of a layer of adsorbed particles on the solid surface due to strong
attractive forces between these dispersed particles and the surface. This layer acts as a `soft
cushion' over which the particles in the fluid slide \cite{SethCB2008}. Sedimentation due to
gravitational or other forces could also be at the origin of this mechanism. A third mechanism for
slip at wall is the formation of a thin lubricating film at the wall due to the presence of
additives which migrate to the boundary region to form the film or to reduce the viscosity of the
mixture in that region \cite{KissiP1996, PitHL1999}. Slip may also be explained by the presence of
a thin film of a gaseous phase on the surface which acts as a lubricant layer \cite{Gennes1985,
TrethewayM2002, NetoCW2003, GranickZL2003, HenryNEBC2004}.

Several other mechanisms have also been proposed to explain wall slip. These include adhesive
failure at the interface, cohesive failure in the vicinity of the interface, chain disentanglement,
de-bonding and desorption events \cite{LukMA1987, HatzikiriakosD1991, ChenKB1993, Graham1995,
BlackG1996, LegerHMD1997, Larsonbook1999, Wang1999, JoshiLM2000, Denn2001, MacielSSGM2002,
DubbeldamM2003, WallsCSK2003, RobertDV2004, GonzalezGSV2009}. Some of these mechanisms are
essentially the same or can originate from other mechanisms. For example adhesive failure can
originate from desorption events and cohesive failure can occur by disentanglement
\cite{WangDI1996, Denn2001, RobertDV2004}.

It is noteworthy that all these slip mechanisms could apply in different flow systems; moreover
some of these mechanisms could occur in the same system simultaneously or under different
deformation regimes and flow conditions \cite{Denn2001}. However, the validity of any proposed
mechanism is solely dependent on observational confirmation in real life experiments. In fact some
of the above mentioned mechanisms, such as stress gradient, have been criticized by the lack of
experimental support.

%SSSSSSSSSSSSSSSSSSSSSSSSSSSSSSSSSSSSS
\section{Slip Signatures}

Wall slip has been observed directly in many experimental studies through the use of various
imaging and visualization techniques (refer to \S\ \ref{Measurement}). However, most often it is
detected indirectly through its effects on the flow \cite{MiglerHL1993, BlackG1996, MeekerBC2004a,
MeekerBC2004b}. There are several signatures that reveal the occurrence of slip in various
rheological and fluid transport systems. One of these signatures is the dependence of the
rheological properties, like viscosity, on the flow geometry such as the gap width in
parallel-plate rheometers or the shape and radius of capillaries \cite{Nguyen1992, YilmazerK1992,
KalyonYAY1993, Barnes1995, GoshawkBKG1998, Larsonbook1999, RusselG2000, Pal2000, JoshiLM2000,
MacielSSGM2002, CheikhK2003, WallsCSK2003, MeekerBC2004a, Bertola2007, MollerFB2009, ChenDZY2009}.
The effect of slip generally increases as the area-to-volume ratio of the flow path increases
\cite{KalyonG2003, Kalyon2005}; hence, slip effects are normally more pronounced in small
capillaries, narrow gap rheometers and porous media \cite{CohenM1985, Kiljanski1989,
FletcherFLLBe1991, HatzikiriakosD1991, Nguyen1992, HollandB1995, Barnes1995, CheikhK2003,
AryaCM2003, Kalyon2005, ChenDZY2009}. This can be explained by increasing the relative importance
of slip contribution to the total volumetric flow rate due to the fact that this contribution is
proportional to the slip area with a quadratic/cubic dependence of area/volume on length. This has
also been explained by proposing that the slip effect has a significant impact only when the length
scale over which the liquid velocity changes is comparable to the slip length \cite{ChenKB1993,
CraigNW2001, NetoCW2003, TrethewayM2004, GovardhanSAB2009}. In disperse systems, this can also be
justified by an increase in the slip significance with an increase in the size of suspended
particles relative to the dimensions of flow paths \cite{Nguyen1992, Pal2000}. Other explanations;
such as increasing the degree of structural destruction which causes the formation of slip layer at
the wall with reducing the size \cite{ChangKS2003}, or the dependence of slip effect on the ratio
of the slip layer width to the dimension of flow channel \cite{MacielSSGM2002}; have also been
proposed.

Another signature of wall slip is the dependence of the rheological and flow properties on the
composition of the surface and its physicochemical characteristics such as wettability and
roughness \cite{Larsonbook1999, WallsCSK2003}. A third signature, which is specific to viscoplastic
materials, is disappearance of yield-stress or a drop in the yield-stress value \cite{SpathisM1997,
PlucinskiGC1998}. A fourth signature is a sudden drop in the shear viscosity or shear stress
associated with flow enhancement at a critical pressure or stress \cite{CohenM1985, CohenM1986,
HatzikiriakosD1991, Barnes1995, BlackG1996, Toorman1997, FrancoGB1998, JoshiLM2000, ChangKS2003,
KalyonG2003}.

The occurrence of slip can also be revealed by the presence of distortions; such as fluctuations,
discontinuities, non-monotonic trends and nonuniform slopes; in the shape of rheograms and other
flow characterization curves \cite{Westover1966, BlylerH1970, MagninP1987, MagninP1990,
HatzikiriakosD1991, MiglerHL1993, Barnes1995, WangDI1996, Larsonbook1999, JoshiLM2000, Denn2001,
MacielSSGM2002, ChangKS2003, KalyonG2003}. Some of these curves, such as viscosity versus shear
stress, can be more slip-revealing than others, such as viscosity versus shear rate, as the former
are more sensitive to slip effects than the latter \cite{Barnes1995, Toorman1997, PlucinskiGC1998,
FrancoGB1998, SaintpereHTJ1999, Pal2000, RobertsBC2001, ChangKS2003, DenkovSGL2005,
DurairajMSE2009}. Failure to reproduce certain results for a particular flow system under similar
physical conditions can strongly indicate the presence of wall slip \cite{NguyenB1985, Nguyen1992,
Barnes1995, SanchezVFG2001}. This may be manifested by the so-called inherent constitutive
instabilities where non-unique relationships do exist, as can be identified from the appearance of
multiple values of the dependent variable against a single value of the independent variable in the
flow characterization curves such as shear stress versus shear rate \cite{Graham1995, BlackG1996,
Larsonbook1999, MacielSSGM2002}. The reason for this multiplicity is that wall slip depends on
minute details that are difficult to control and regenerate especially when these details belong to
the microscopic world \cite{NetoEBBC2005}. An example that demonstrates this reality is what has
been observed by some investigators \cite{BizonnJBCRC2002} of ``a significant variability of the
slip lengths measured on different hydrophobic samples prepared with the same experimental
procedure''. This fact may partially explain the contradicting experimental results obtained from
very similar fluid systems by different investigators on the effect of factors; such as
wettability, roughness and deformation rate; on wall slip \cite{NetoCW2003, BizonneCSC2005,
HenryNEBC2004, NetoEBBC2005}. These results vary not only quantitatively, and some times grossly,
but even qualitatively. The difference in instrumentations, techniques and interpretations, as well
as experimental and theoretical errors, can only offer a partial explanation to these
contradictions. In some cases, the contradicting results may also be explained by the involvement
of other flow phenomena, such as shear banding, whose effects resemble and hence confuse the effect
of wall slip.

%SSSSSSSSSSSSSSSSSSSSSSSSSSSSSSSSSSSSS
\section{Slip Measurement} \label{Measurement}

The traditional method for detecting and measuring wall slip is by the use of Mooney plots
\cite{BlylerH1970, ChenKB1992, RosenbaumH1997, MacielSSGM2002} rather than by direct observation
and measurement as it is commonly done in the recent years \cite{ChangKS2003}. Mooney method
\cite{Mooney1931} is based on the use of flow curves, i.e. shear stress at wall against apparent
shear rate, from at least three capillaries \cite{Kiljanski1989, MacielSSGM2002} of different size
but with the same length-to-diameter ratio so that the effect of pressure drop and diameter size
remains constant \cite{RosenbaumH1997, LamWCJ2007}. The dependence of flow curves on diameter then
indicates the presence of slip \cite{YilmazerK1992}. The shear stress at which this dependence
emerges is taken as the critical value for the onset of slip. Because the slip velocity is
correlated to the variation of the velocity gradient as a function of the diameter size for a
certain shear stress, the slip velocity can be evaluated from this variation. By plotting the
apparent wall shear rate versus the reciprocal capillary diameter for a given wall shear stress, a
straight line should be obtained. The slip velocity can then be extracted from the slope (slope = 8
$\times$ velocity) and the no-slip (true) shear rate from the $y$-intercept \cite{CohenM1985,
ChenKB1992, ChenKB1993, KalyonYAY1993, Barnes1995, HatzikiriakosHHS1995}. The slip velocity values
obtained from this procedure can then be used to find the slip corrections required to obtain the
slip-free correlation between shear rate and shear stress \cite{CohenM1985}. A similar procedure is
followed to obtain the slip parameters from other rheometric geometries, such as parallel-plate,
where the gap width replaces the capillary diameter with a different slop scale
\cite{YilmazerK1992, Larsonbook1999, EggerM2006}.

However, Mooney method is based on an approximate analysis with an implicit assumption that slip
velocity depends only on wall shear stress \cite{ChenDZY2009}. Hence, it ignores the effect of
pressure, temperature and viscous heating which could be particularly important at high flow rates
and in cases of pressure-dependent slip in the presence of large pressure gradients
\cite{HatzikiriakosHHS1995, RosenbaumH1997, SalmonBMC2003}. It also ignores the end effects and
implicitly assumes a fully developed flow of homogeneous bulk material which may not be true in
general \cite{ChenKB1993, LamWCJ2007}. Furthermore it may be inconvenient as it requires the use of
multiple geometries in a number of measurements \cite{Kiljanski1989, AralK1994, ChakrabandhuS2005},
and involve various sources of approximation errors in the measurements and subsequent
calculations. For example, several investigators (e.g. \cite{HatzikiriakosHHS1995}) have pointed
out to the fact that the assumption of linear relationship between the reciprocal diameter and
apparent shear rate, which Mooney method is based upon \cite{KalyonYAY1993}, is not valid in some
practical situations \cite{RosenbaumH1997, MacielSSGM2002}.

For such reasons, a number of slip models have been proposed to replace or improve Mooney method
(see for example \cite{YoshimuraP1988, Kiljanski1989, HatzikiriakosD1992, YeowLMM2003, YeowCKS2004,
MeekerBC2004a, ChakrabandhuS2005, ChenDZY2009}). Some of these models (e.g. \cite{YoshimuraP1988,
Kiljanski1989, YeowCKS2004}) require only two data sets from two rheometric measurements with
relaxed restrictions on the dimensions of rheometric apparatus. Other modifications to Mooney
method have also been proposed to accommodate the dependence of slip velocity on factors other than
wall shear stress such as the radius of capillary \cite{HollandB1995, ChenDZY2009}. All these
proposals, like Mooney method, are empirical models and hence are not based on a fundamental
physical theory that provides intuitive insight \cite{CohenM1985}. Some investigators (e.g.
\cite{AstaritaMBook1974, BonillaS2000}) use a minimum of two flow data sets (shear stress versus
shear rate) from tubes of different diameters to detect the onset of slip. The absence of slip is
then indicated by the coincidence of flow curves.

A number of direct observation and measurement techniques have been developed, mainly in the
context of rheological studies and rheometry, to monitor slip and quantify its velocity or to
measure the thickness of slip layer and extrapolation length. These techniques include the use of
marker lines \cite{Princen1985, MagninP1987, KalyonYAY1993, AralK1994, LawalK1997, PlucinskiGC1998,
KalyonG2003}, optical photography \cite{MagninP1990, BoersmaBLS1991}, evanescent-wave-induced
fluorescence combined with fringe pattern fluorescence recovery \cite{MiglerHL1993}, nuclear
magnetic resonance (NMR) imaging \cite{RofeLCVG1996, CallaghanCRS1996, BrittonC1997}, near field
laser velocimetry \cite{LegerHMD1997, HervetL2003, SchmatkoHL2005}, total internal
reflection-fluorescence recovery \cite{PitHL1999, PitHL2000}, high-speed cinematography
\cite{GevgililiK2001}, surface force apparatus \cite{BaudryCTM2001, GranickZL2003}, infrared
spectroscopy \cite{HartmanKokKLB2002}, particle image velocimetry \cite{TrethewayM2002,
TrethewayM2003, JosephT2005, GonzalezGSV2009}, scanning electron microscopy \cite{MacielSSGM2002,
KalyonG2003}, heterodyne dynamic light scattering \cite{SalmonBMC2003}, atomic force microscopy
\cite{BonaccursoBC2003, NetoCW2003, ChoLR2004, HenryNEBC2004}, video microscopy
\cite{MeekerBC2004a, MeekerBC2004b, SethCB2008}, laser Doppler velocimetry \cite{RobertDV2004},
total internal reflection velocimetry \cite{HuangGB2006, HuangB2007}, particle tracking velocimetry
\cite{BoukanyW2009}, and ultrasonic speckle velocimetry \cite{LettingaM2009}.

Some of these techniques may rely in their conclusions on tentative theoretical models or suffer
from limited spatial resolution \cite{Denn2001, Kalyon2005, LamWCJ2007}. As the slip velocity can
be obtained directly by some of these techniques, the use of analytical and empirical relations to
deduce slip velocity becomes redundant \cite{KalyonYAY1993}. Another advantage of these techniques
is that they can be used to detect and characterize temperature- and time-dependent quantities,
such as shear viscosity and slip velocity during transient states, as well as steady-state
quantities \cite{KalyonYAY1993, AralK1994}. Visual inspection can also be used in some situations
to detect slip especially in yield-stress fluid systems where the bulk remains solid and wall slip
can be easily observed \cite{MagninP1990, PlucinskiGC1998, LindnerCB2000}.

Molecular dynamics simulations have been used extensively \cite{ThompsonR1990, SunE1992,
MorrisHG1992, GuptaCC1997, JabbarzadehAT1999, BarratB1999, CieplakKB2001, AryaCM2003,
VoronovPL2007, LichterMSW2007} to investigate slip where direct experimental observation is hard or
impossible to achieve \cite{ZhangK2004}, or where the use of computational tools is more
convenient, or for the purpose of experimental validation and comparison \cite{AryaCM2003}. The
advantage of molecular dynamics approach is that it is capable of dealing with extreme system
conditions such as very high shear rates \cite{JabbarzadehAT1999}, short time scales and nanoscopic
dimensions where the system behavior is dominated by molecular scale processes. Other simulation
and numerical methods have also been employed to investigate slip and slip-related issues
\cite{BillerP1987, Sorbie19902, MorrisHG1992, Hatzikiriakos1994, RosenbaumH1997, PitHL1999,
JoshiLM2000, DubbeldamM2003, Georgiou2003, RobertDV2004, JiYC2006, SaugeyDW2006, HartingKH2006,
YeowLK2006, KalyonT2007, LamWCJ2007, YanHB2008, ShamsKH2009}. Some of these methods have been used
to overcome stated limitations in molecular dynamics (MD) simulations of not being able to reach
low shear rates and large system sizes \cite{HartingKH2006} because MD simulations are usually
limited in size to systems on molecular scale (or a few multiples larger) due to practical
restrictions on the computational resources. It is noteworthy that nonslip condition may be relaxed
in some numerical simulations to avoid singularities \cite{PozrikidisBook2001}.

%SSSSSSSSSSSSSSSSSSSSSSSSSSSSSSSSSSSSS
\section{Slip in Rheometry}

Most studies on wall slip have been carried out in the context of rheology and rheometric
measurements rather than in flow transport systems. This is highlighted by the fact that most slip
studies come from the literature of rheology. One reason is that the rheometric apparatus are
better equipped to detect and quantify this phenomenon. Another reason is that slip is commonplace
in rheometry \cite{Nguyen1992}, as it has been observed in various types of rheometric systems, and
is relatively easy to detect within the rheometric settings by different techniques including
direct observation. It is rather obvious that detecting and measuring slip in a parallel-plate
rheometer, for example, is easier than in a porous medium.

Wall slip is a major source of error in rheometry. When slip occurs, the apparent shear rate, as
measured from deformation speed and system geometry, is different from the real shear rate because
the apparent shear rate combines the effect of fluid deformation and wall slip \cite{KalyonYAY1993,
MeekerBC2004a}. Hence, wall slip can introduce gross errors on rheological measurements rendering
them invalid \cite{Westover1966, Princen1985, MagninP1990, GoshawkBKG1998, SaakJS2001, ChangKS2003,
KelessidisM2008, ChenDZY2009}. Corrections, which normally require considerable experimental and
theoretical effort, should therefore be applied if slip cannot be prevented \cite{YilmazerK1992,
KalyonT2007}. Since wall slip generally increases with increasing surface-to-volume ratio
\cite{KalyonYAY1993, TrethewayM2004}, accounting for slip is particularly important in rheometry
because of the common use of small tubes and narrow gaps \cite{Kiljanski1989, ChenKB1993,
GoshawkBKG1998}.

Slip in rheometry is manifested by the dependence of the measured values, such as shear viscosity
and elastic modulus, on the dimensions and shape of the measuring device. To verify that wall slip
has negligible impact on the rheological measurements, rheometers of different type or size may be
used simultaneously and the results are compared. The dependence of the measurements on the type or
size will be a strong sign on the occurrence of wall slip \cite{CohenM1985, PlucinskiGC1998,
SaintpereHTJ1999, WallsCSK2003}.

Rheometric geometries vary in their vulnerability to slip. For example, the cone-and-plate and
parallel-plate geometries are normally susceptible to slip \cite{Barnes1995, FrancoGB1998}. The
vane geometry, on the other hand, avoids slip and hence is widely used in rheological measurements
in general and those related to yield-stress fluids in particular \cite{NguyenB1985, JamesWW1987,
Nguyen1992, Barnes1995, LiddellB1996, Barnes1999, BarnesN2001, RobertsBC2001, SaakJS2001,
HarteCC2007, KelessidisM2008}. Another advantage of the vane geometry is that it involves less
structural disruption which is particularly important for the measurements of shear-sensitive and
thixotropic/rheopectic fluids \cite{JamesWW1987, Nguyen1992, LiddellB1996, BarnesN2001,
HarteCC2007, KelessidisM2008}. However, a disadvantage of the vane geometry is the requirement of a
larger sample volume compared to some other geometries such as capillary and parallel-plate
\cite{WallsCSK2003}. Because wall slip is not taken into consideration in the design and
manufacture of common types of commercial rheometers, careful and elaborate experimental protocols
are usually followed to eliminate, reduce or detect and quantify slip. Some researchers have even
built their own rheometers to avoid wall slip and make more precise measurements
\cite{LiddellB1996}.

The effects of slip could be particularly significant in some rheometric measurements such as those
involving the application of step strain, as used for example in measuring the stress relaxation
modulus of viscoelastic fluids \cite{ArcherCL1995, GevgililiK2001, VenerusN2006, Ottinger2008,
SochiVE2009, SochiFeature2010}. This is due to an increased importance of other effects such as
inertia, time-dependency and elasticity.

%SSSSSSSSSSSSSSSSSSSSSSSSSSSSSSSSSSSSS
\section{Slip in Porous Media}

Although wall slip has been discussed in a number of studies within the context of flow through
porous media and packed beds \cite{KozickiHT1967, MarshallM1967, Chauveteau1982, LukMA1987,
FletcherFLLBe1991, HuangS1992} and has been suggested as a possible reason for some observed
anomalies, it has not been given the consideration that it deserves. The obvious reason is that
wall slip is very hard to observe and measure directly in the porous media \cite{Sorbie1990}, and
hence most results are based on deduction, theoretical analysis and computer simulation
\cite{Sorbie1990, Sorbie19902}. In fact, some of these discussions are little more than
speculations and general statements. A principal indicator that is commonly used to detect and
quantify slip in porous media is the occurrence of a tangible drop in the friction factor as
predicted by Ergun's equation \cite{LukMA1987}. The effect of slip on the flow of gases through
porous media is particularly important due to the fact that the mean free path of the gas molecules
could be comparable to the pore size \cite{Dullien1975, TiltonBook2007}. In micro- and nano-porous
media, theoretical and experimental evidence indicate that significant wall slip could occur
resulting in a considerable increase in the effective permeability \cite{CraigNW2001,
TrethewayM2002, TrethewayM2004}. The effect of slip in porous media may be considered by
introducing the Klinkenberg permeability correction to Darcy's law \cite{Dullien1975}.

The occurrence of slip in the flow through porous media has been asserted in a number of
experimental studies by indirect observation \cite{Chauveteau1982, LukMA1987, FletcherFLLBe1991,
HuangS1992, Barnes1995} using a variety of fluids and core materials under various physical
conditions. Some of these studies have reported the observation of very large and clear slip
effects in certain flow systems. As pointed out already, one of the slip signatures is the
dependence of rheological and flow parameters; such as viscosity, yield-stress and volumetric flow
rate; on the size and shape of the flow paths. This can be manifested in the observed deviation of
the rheological and flow properties in porous media from the predictions as obtained from
theoretical analysis, slip-corrected measurements and large duct flow \cite{CohenM1985, CohenM1986,
SorbieCJ1989, Sorbie1990, Sorbie19902, Barnes1995, LiuM1998}. For example, the decline in apparent
viscosity, as observed in some porous media flow experiments, compared to the bulk viscosity can be
explained by slip at the pore walls \cite{FletcherFLLBe1991, HuangS1992}, although other factors,
such as amplification of shear-rate-dependent effects, may be behind, or at least contribute to,
this phenomenon. The effect of slip seems to be more pronounced in lower permeability porous media
within certain restrictions \cite{Sorbie1990}. This is due to the effect of increasing the
area-to-volume ratio on reducing the size of the flow paths because of the general relation between
a lower permeability and smaller pores.

%SSSSSSSSSSSS
\section{Slip Quantification}

The slip at wall is quantified by the relative fluid-solid velocity at the interface, e.g. 1.5
mm/s. An approximate estimate of the slip velocity may be obtained from a modified form of Eyring's
equation for the shear rate across a layer of molecules \cite{BlackG1996, Larsonbook1999}. As
pointed out earlier, slip velocity is a function of the wall shear stress \cite{KozickiCT1966,
Hatzikiriakos1994, SalmonBMC2003}. The thickness of the slip layer is also a function of the wall
shear stress \cite{KalyonYAY1993, MacielSSGM2002, SalmonBMC2003}. The correlation between the slip
velocity and wall shear stress, which in many cases is assumed to be linear although it may be
non-monotonic with deformation rate variation \cite{Larsonbook1999, DubbeldamM2003}, is linked
through the slip coefficient \cite{Mooney1931, CohenM1985, YilmazerK1989, ChenKB1992,
HollandB1995}. The slip velocity can also be characterized by the shear rate at wall with the use
of a slip length proportionality factor, as given in \S\ \ref{Mechanisms}.

Since slip velocity depends on several factors, such as fluid and surface properties as well as the
surrounding conditions like pressure and temperature \cite{KozickiHT1967, HatzikiriakosD1991,
Hatzikiriakos1994, ChakrabandhuS2005}, the slip coefficient is system-dependent \cite{BerkerV1992,
LawalK1997, LawalK1998}. This dependency allows the above-mentioned simplification of the relation
between slip velocity and wall shear stress as a direct proportionality, where the other factors
are absorbed in the slip coefficient \cite{LukMA1987, Vinogradova1999}. The slip coefficient could
also be a function of the wall shear stress, possibly over a certain range of deformation rate
\cite{BerkerV1992, KalyonYAY1993, BonillaS2000, Kalyon2005, ChakrabandhuS2005}. The slip
coefficient may be affected by other factors such as Reynolds number and static contact angle for
wetting fluids \cite{LukMA1987}.

Nonlinear types of slip velocity dependence on wall shear stress, such as quadratic
\cite{SalmonBMC2003, MeekerBC2004a, YeowLK2006, SethCB2008}, cubic \cite{Barnes1995} and power law
\cite{CohenM1985, HatzikiriakosD1991, Hatzikiriakos1994, Barnes1995, BlackG1996, KumarG1998,
MacielSSGM2002, ChakrabandhuS2005, VieraDFG2006}, have also been proposed in the literature
\cite{Kalyon2005, LamWCJ2007, ChenDZY2009}. In the case of power law dependence, which apparently
is the most common form \cite{MacielSSGM2002}, the exponent of the power law, as well as its
coefficient, usually vary depending on the fluid-solid system and ambient conditions
\cite{CohenM1985, Hatzikiriakos1994}. Similar sort of variation could also be assumed for other
types of correlation. Different types of correlation may apply to the same system over different
deformation rate regimes and under different physical conditions \cite{Hatzikiriakos1994,
LegerHMD1997, SalmonBMC2003}. Most of these correlations represent reasonable approximations for
the real system behavior under practical situations, although this behavior is usually more complex
to be expressed by a simple analytical form \cite{SalmonBMC2003}.

The thickness of the slip layer varies depending on the fluid-solid system characteristics (such as
viscosity, elastic modulus, bulk density, particle size, concentration, surface composition,
surface roughness, and flow geometry) as well as the deformation rate and surrounding conditions
\cite{YilmazerK1992, KalyonYAY1993, AralK1994, Barnes1995, SaakJS2001, MacielSSGM2002, Kalyon2005,
EggerM2006, HartingKH2006, LamWCJ2007, ChenDZY2009}. Most values reported in the literature give an
estimate of the slip layer thickness in the order of a micron \cite{YilmazerK1989, YilmazerK1992,
KalyonYAY1993, SaakJS2001, HartmanKokKLB2002, MacielSSGM2002, Kalyon2005, EggerM2006, VieraDFG2006,
TropeaYF2007, LamWCJ2007} with some exceptionally low values in the nanometer range
\cite{GoshawkBKG1998, RusselG2000, TrethewayM2003, MeekerBC2004a, MeekerBC2004b}. Some large values
in the millimeter regime have also been reported with a justification that the thickness of slip
layer is of the order of the dimension of suspended particles \cite{ChakrabandhuS2005}. The slip
layer thickness also depends on the nature of the layer, such as being a depleted layer from the
continuum phase or consisting of another phase like a gas film on a liquid-solid interface. The
thickness could be time-dependent during transitional flow regimes before reaching a final
equilibrium value at a steady-state flow \cite{KalyonYAY1993, AralK1994}.

%SSSSSSSSSSSSSSSSSSSSSSSSSSSSSSSSSSSSS
\section{Slip Instabilities}

As indicated already, wall slip has been blamed for inducing or encouraging flow instabilities and
melt fracture or at least enhancing the propagation of disturbances which causes these
instabilities, although slip is also believed to bring stability to some systems \cite{BlackG1996,
Denn2001, KalyonG2003}. Slip can be periodic where consecutive cycles of slip and stick occur, the
so-called `stick-slip' or `spurt flow' phenomenon which has been observed in several types of fluid
system \cite{Westover1966, KaoNH1975, Denn1990, Barnes1995, Larsonbook1999, Wang1999, Denn2001,
Georgiou2003, RobertDV2004, GonzalezGSV2009}. Stick-slip seems to initiate at a critical,
system-dependent shear stress and lasts for a limited range of deformation rate \cite{WangDI1996,
KumarG1998, Wang1999, Denn2001, HervetL2003, KalyonG2003, RobertDV2004}. The critical stress can be
at the origin of this alternating cycle where nonslip condition fails on exceeding the threshold
stress, followed by the establishment of stick condition, as the stress falls below the critical
value, with a subsequent stress buildup \cite{RobertDV2004}. The occurrence, duration and critical
stress of stick-slip are influenced by the characteristics of fluid-solid system and the
surrounding conditions \cite{Schowalter1988, Denn2001}. The stick-slip phenomenon, which is
commonplace in solid friction systems, is related to instabilities and oscillating time-dependent
effects in the flow field \cite{WangDI1996, Wang1999, DubbeldamM2003}, where thixotropy/rheopexy
and viscoelasticity can be at the origin of these effects \cite{Westover1966, HatzikiriakosD1991,
Graham1995, BlackG1996, Denn2001}. It may also be related to structural turbulence where viscous
and elastic forces at the interface compete and hence play alternating roles \cite{MccombBook1992}.

In some cases, the stick-slip behavior is associated with a process of alternating fracture-reheal
of the sheared sample \cite{HuB1998, PicardABL2002, GonzalezGSV2009}. Stick-slip appears to be more
common in yield-stress fluid systems at relatively low deformation regimes \cite{LiddellB1996,
PicardABL2002}. Several explanations have been proposed for the origin of slip during stick-slip
occurrence. These include adhesive or cohesive failure and chain disentanglement mechanisms
\cite{Denn2001, DubbeldamM2003, RobertDV2004, GonzalezGSV2009}. Slip in general, including that in
the stick-slip instability, can also occur through other mechanisms as discussed in \S\
\ref{Mechanisms}. Most of these mechanisms can apply during stick-slip instability in different
systems and under different circumstances. Stick-slip occurrence is clearly manifested by the
appearance of periodic oscillations on the flow curves such as flow rate versus applied pressure.

Wall slip may also be blamed for inducing the formation of sharkskin melt fracture in polymeric
systems, or at least being one of the possible causes or a contributing factor to this phenomenon
\cite{Hatzikiriakos1994, HatzikiriakosHHS1995, Graham1995, WangDI1996, BlackG1996, Wang1999,
Denn2001}. Sharkskin fracture could be related to locally varying slip at certain interface regions
due possibly to inhomogeneous surface or fluid conditions. The onset of sharkskin seems to occur at
a critical shear rate and lasts over a limited range of deformation rate \cite{Hatzikiriakos1994,
HatzikiriakosHHS1995, WangDI1996, Denn2001, KalyonG2003, RobertDV2004}. The critical shear stress
for the onset of sharkskin is usually higher than that for the onset of slip
\cite{Hatzikiriakos1994}. This implies that even if the slip is a necessary condition for the
sharkskin formation, it is not a sufficient one. Other factors; such as sample acceleration at the
exit region, flow rate, temperature and extensional effects; could also contribute to the onset and
severity of sharkskin defect \cite{Hatzikiriakos1994, HatzikiriakosHHS1995}. Sharkskin may also
originate from, or associate, stick-slip instability where the oscillating fluctuations at the
interface affect the texture of the melt surface, leading to the sharkskin distortion
\cite{Hatzikiriakos1994, WangDI1996, Georgiou2003}. Oscillating behavior between sharkskin and
smooth surface have also been observed during the stick-slip instability \cite{Denn2001,
KalyonG2003}. This may point out to the involvement of two different instability wavelengths.
Sharkskin seems to be highly dependent on the material of the polymer melt, as well as its
dependency on the material and properties of the surface and the external conditions
\cite{Denn2001}. It should be remarked that flow instabilities associated with wall slip can be
identified from the trend and turning points on the flow characterization curves
\cite{KalyonG2003}.

In conclusion, although wall slip seems to play a role in at least some of these flow
instabilities, the exact nature of this role is not well understood \cite{HatzikiriakosHHS1995}.
Moreover, this role may be different in different fluid-solid systems, physical conditions and
types of instability.

%SSSSSSSSSSSSSSSSSSSSSSSSSSSSSSSSSSSSS
\section{Slip and non-Newtonian Behavior}

Wall slip occurs in Newtonian as well as non-Newtonian fluid systems \cite{Gennes1985, ZhuG2001,
CraigNW2001, NetoCW2003, GranickZL2003, BonaccursoBC2003, HenryNEBC2004}, although its occurrence
in non-Newtonian is more common than in Newtonian \cite{KozickiCT1966, HatzikiriakosD1991,
ChangKS2003}. The reason is that the complexity of non-Newtonian fluids, especially multi-component
fluids, allows the introduction of many factors that promote this phenomenon. Slip seems to be the
exception in Newtonian systems and the norm in, at least, some types of non-Newtonian systems such
as dispersions and yield-stress fluids \cite{Schowalter1988, DenkovSGL2005, LettingaM2009,
Buscall2010}. The effects of slip are particularly important in non-Newtonian systems as the steep
velocity gradients at the boundary layer means high shear rates with subsequent amplification of
shear-rate-induced effects such as shear-thinning, shear-thickening and time-dependency
\cite{MacielSSGM2002}.

Experimental observations indicate that the slip contribution to the flow rate increases with
increasing wall shear stress in the shear-thickening fluids, while it decreases in the
shear-thinning fluids \cite{CohenM1985, YilmazerK1992, Barnes1995, ChenDZY2009}. Although the
general trend in wall slip is that it normally results in an apparent drop in the bulk viscosity
\cite{Nguyen1992, SanchezVFG2001} and hence in an increase in the flow throughput, in the case of
shear-thinning fluids the slip could, in principle, result in an increase in the apparent viscosity
and a drop in the flow throughput if the extra flow due to slip does not compensate for the loss of
flow due to the lack of deformation and consequent shear-thinning. Similar argument may also apply
to yield-stress fluids in the case of partial yield with a mixed plug and bulk flow due to
slip-related inhomogeneous stress distribution.

Wall slip could be behind some of the observed plateaux in the flow characterization curves, such
as rheograms, where a flow parameter remains constant over a certain range of deformation rate.
These plateaux have been observed in many fluid systems during various deformation rate regimes
\cite{MagninP1990}. For example, the intermediate viscosity plateau at intermediate shear rates, as
seen in some viscoelastic systems \cite{SochiVE2009, SochiArticle2010}, could in some cases
originate from wall slip \cite{FrancoGB1998} which causes a sudden sharp drop in viscosity then
holds the viscosity virtually constant over a certain range of shear rate. Another possibility for
the appearance of this plateau is that dynamic yield-stress is responsible for the sudden drop in
viscosity while wall slip is responsible for the flat plateau \cite{RusselG2000}. Other
explanations, not related to slip, for the appearance of these intermediate plateaux in
viscoelastic systems have also been proposed \cite{SochiVE2009, SochiArticle2010}. Another example
of slip-related plateaux is the Newtonian plateau at very low shear rates, as observed for instance
in some yield-stress fluids \cite{Barnes1995, WallsCSK2003, Meeten2004}, which hides the
viscoplastic nature of these materials, as will be discussed in \S\ \ref{Yield}.

In the following subsections we discuss some issues related to wall slip with regard to certain
types of non-Newtonian behavior.

%SSSSSSSSSSSSSSSSSSSSSSSSSSSSSSSSSSSSS
\subsection{Yield-Stress} \label{Yield}

Wall slip is commonplace in yield-stress fluid systems. One consequence of this is that there is
normally an apparent yield due to slip at lower stress than the bulk yield-stress. This could
obscure the viscoplastic nature of the material because the apparent flow due to slip hides the
yield-stress behavior \cite{KalyonYAY1993, SaintpereHTJ1999, Larsonbook1999, RusselG2000,
LindnerCB2000, WallsCSK2003}. In these cases, direct observation techniques, such as visualization,
may be required to clarify the situation \cite{MagninP1990, KalyonYAY1993}. Slip could also result
in a misleading lower value of yield-stress because although the material appears to flow at a
lower value of stress it does so by sliding rather than by undergoing bulk deformation
\cite{Westover1966, JamesWW1987, SpathisM1997, PlucinskiGC1998, WallsCSK2003, SethCB2008}. Because
the local stress is below the yield value except possibly at the boundary region, no yielding
occurs except at a very thin layer adjacent to the surface, which serves as a lubricating slip
film. There is also the possibility that the sliding occurs through a depleted inhomogeneous layer
from the continuum phase, or through a film from another phase which exists on the surface prior to
the contact or is deposited on the surface from the bulk phase \cite{Princen1985, Nguyen1992,
Larsonbook1999}. The lower yield-stress value due to slip may be called `sliding' or `apparent'
yield-stress \cite{MeekerBC2004a, MeekerBC2004b, SethCB2008}. Slip can also lead to large errors in
viscosity estimations and flow rate predictions in these systems \cite{Westover1966, MagninP1990,
KalyonYAY1993, PlucinskiGC1998, SaintpereHTJ1999, SethCB2008}. It should be remarked that slip
could also happen at high deformation rate regimes following bulk-yielding in association with the
bulk flow \cite{KaoNH1975, JamesWW1987, SaintpereHTJ1999, LindnerCB2000, SethCB2008}.

The presence of slip can also lead to a false yield-stress in non-viscoplastic fluids, where a
nonslip at a low deformation regime followed by a sudden slip at a higher deformation regime can
give a misleading impression of yielding, which is manifested in a sudden dip in rheograms
\cite{Mannheimer1991, Barnes1995, FrancoGB1998, Barnes1999, RusselG2000}. The dependence of this
behavior on factors, such as rheometric geometry and surface conditions, would reveal its origin as
arising from slip rather than viscoplastic yield. Wall slip may also be responsible for the large
differences between the static and dynamic yield-stress measurements as obtained from some
rheometric geometries \cite{SaakJS2001, KelessidisM2008}, although other explanations have also
been proposed to explain these discrepancies \cite{SochiB2008, SochiYield2010}.

In an experimental study on the flow of highly concentrated suspensions of soft particles, it has
been reported \cite{MeekerBC2004a, MeekerBC2004b} that depending on the value of the applied stress
three regimes of slip have been detected: at low stress (at and below yield-stress) the flow is
entirely due to wall slip, slightly above the yield-stress both bulk flow and slip have significant
contribution to the fluid motion, and well above yield-stress the slip is negligible and the flow
is almost entirely due to bulk deformation. Similar observations have been reported in other
studies \cite{Nguyen1992, SethCB2008}. Although this trend of behavior is quite logical and may be
valid for other types of yield-stress systems, experimental verification is required to generalize
these results to other systems.

It is noteworthy that wall slip may be exploited beneficially to measure the yield-stress of
viscoplastic materials \cite{Nguyen1992}. Real yield in viscoplastic materials usually marks the
transition from plug flow to bulk flow where deformation-induced flow dominates and the
significance of slip contribution to the total flow rate reduces dramatically \cite{Princen1985,
ChenDZY2009}. This transition can be used to estimate the yield-stress value of viscoplastic
materials \cite{Kalyon2005, ChenDZY2009}.

%SSSSSSSSSSSSSSSSSSSSSSSSSSSSSSSSSSSSS
\subsection{Viscoelasticity}

As indicated in \S\ \ref{Factors}, there is a correlation between wall slip and the elastic
properties of fluid as characterized by the elastic modulus. In some types of concentrated
dispersed systems, the slip velocity may vary linearly with the elastic modulus
\cite{MeekerBC2004a, MeekerBC2004b, SethCB2006}. In general, the elasticity of fluid has a
significant impact on the occurrence and magnitude of wall slip \cite{Graham1995}. This fact is
demonstrated more vividly by some proposed mechanisms for wall slip and the formation of slip layer
(see for example \cite{MeekerBC2004a} on the elasto-hydrodynamic theory and refer to \S\
\ref{Mechanisms} for further discussion). The elastic properties of the surface also have an impact
on the slip especially when the fluid itself is characterized by a strong elastic nature
\cite{Vinogradova1999}. Theoretical models supported by experimental evidence indicate that slip
can lead to flow instabilities and complex time-dependent behavior in viscoelastic shear flow;
moreover, viscoelastic effects may also be at the origin of some slip-related instabilities
\cite{Graham1995, BlackG1996}. It has been claimed \cite{KalyonG2003} that the observed overshoots
in some viscoelastic systems could be an artifact of wall slip. This view challenges the
generally-accepted constitutive characterization of this phenomenon and may also be
counterintuitive \cite{SochiVE2009, SochiArticle2010}.

With regards to the interaction between wall slip and the extensional flow (which is one of the
characteristic features of viscoelastic fluids), converging-diverging flow channels and other
geometric irregularities (which are particularly important in the flow of viscoelastic fluids
through porous media) \cite{SochiVE2009, SochiFeature2010}, there are few explicit discussions in
the literature of wall slip on these subjects. Extensional flow effects are generally ignored or
given minor consideration in the literature of fluids. It has been asserted \cite{TropeaYF2007}
that when the no-slip at wall condition holds, the flow will be dominated by shear effects, while
when this condition is violated, elongational effects will dominate. However, elongational flow
effects could dominate even when the no-slip condition holds, due for instance to the geometry of
the flow path \cite{SochiVE2009, SochiArticle2010}. One example is the nonslip flow of viscoelastic
fluids in converging-diverging geometries. However, in general slip at wall reduces shear
deformation and encourages extensional flow although other factors still have a significant role in
determining the type of deformation \cite{LawalK1998}. There seem to be a suggestion in
\cite{PlucinskiGC1998} that extensional flow is less susceptible to wall slip, due possibly to the
nature of the measurement technique used in this study. Further discussions concerning the relation
between wall slip and extensional flow could be found in \cite{Larson1988, BarnesbookHW1993,
Hatzikiriakos1994, PlucinskiGC1998}.

%SSSSSSSSSSSSSSSSSSSSSSSSSSSSSSSSSSSSS
\subsection{Time-Dependency}

As indicated already, slip at wall can have time-dependency, as well as space-dependency, due
possibly to temporal variation in the flow field, or to flow instabilities such as stick-slip, or
to the involvement of non-Newtonian effects such as viscoelasticity and thixotropy/rheopexy
\cite{HatzikiriakosD1991, AralK1994, KalyonG2003}. Wall slip can develop through an evolving
structural change in the slip layer at the boundary, or through a dynamic evolution of the depleted
layer which takes time to form during transitional state flow \cite{Barnes1995}. Because these
developments are manifested in a time-dependent viscosity which characterizes thixotropy, wall slip
can be confused with thixotropy \cite{Barnes1995, Barnes1997}. Therefore some of the observed
apparent thixotropic effects could originate from dynamic time-dependent slippage at wall
\cite{BoersmaBLS1991, PlucinskiGC1998}. However, some of these effects can be real thixotropic
effects when they arise from structural changes even if these changes are developed through a wall
slip mechanism and occur mainly at the boundary layer, although they can also be attributed to wall
slip. One factor that can exacerbate the confusion between thixotropy and wall slip is the strong
association between thixotropy and yield-stress fluids which wall slip is one of their
characteristic features \cite{AralK1994, MerkakJM2006, JossicM2009}.

In tube viscometers, time-dependent effects from thixotropic fluids may be detected qualitatively
in some measurements. Because the effect of tube diameter variation caused by the slip of
time-independent fluids is similar to that caused by the flow of time-dependent fluids, certain
procedures have been proposed to distinguish the two effects by varying the tube length with a
fixed diameter and obtaining certain characterization curves. Separate curves will be obtained from
time-dependent fluids, while identical curves will be obtained from slipping time-independent
fluids \cite{Skellandbook1967}.

%SSSSSSSSSSSSSSSSSSSSSSSSSSSSSSSSSSSSS
\section{Slip and Friction}

No-slip condition arises, according to some proposed mechanisms as indicated earlier, because of
static friction between the fluid and the surface. What about the dynamic friction during
fluid-solid interaction when wall slip occurs? Although slip at wall could be non-frictional as
well as frictional depending on the fluid-solid system and ambient conditions \cite{LuptonR1965,
Barnes1995, Wapperomthesis1996, PitHL1999, HervetL2003, DenkovSGL2005, LichterMSW2007,
GonzalezGSV2009}, total frictionless slip may be a remote possibility for viscous flow.
Frictionless slip may be possible for rarefied gas systems if elastic collision between the gas
molecules and the solid is assumed. Anyway, the frictional nature of wall slip should be strongly
related to the type of slip, i.e. true or apparent, and to the slip mechanisms \cite{PitHL1999}.
For example, the depleted layer mechanism should require, at least in most cases, a certain degree
of frictional losses due to viscous dissipation in this layer \cite{OertelBook2004}. In the case of
frictional slip, the strength of frictional forces and the amount of frictional losses should
depend on the fluid-solid system and the ambient physical conditions such as pressure and
temperature \cite{Barnes1995}. As in the case of solid friction, surface roughness normally
increases frictional resistance and frictional losses especially in the case of turbulent flows
\cite{WhiteBook2002}.

%SSSSSSSSSSSSSSSSSSSSSSSSSSSSSSSSSSSSS
\section{Slip Control}

In practical situations where wall slip is unwanted, various methods are employed to eliminate or
minimize slip, although total elimination could sometimes be very difficult or even impossible to
achieve \cite{ChangKS2003}. For example, surfaces are deliberately roughened to reduce the effect
of slip \cite{Barnes1995, LawalK1997, Larsonbook1999, Meeten2004, SethCB2006}. This is achieved by
various techniques such as serrating or coating or impregnating the surface, abrasive blasting,
chemical treatment, and attaching solvent-proof sandpapers \cite{MagninP1990, ChenKB1992,
Nguyen1992, KalyonYAY1993, ChenKB1993, AralK1994, Barnes1995, PlucinskiGC1998, SaintpereHTJ1999,
LindnerCB2000, Pal2000, CiterneCM2001, ChangKS2003, WallsCSK2003, DenkovSGL2005, SethCB2008,
BoukanyW2009}. Some of these techniques could disturb the system by introducing or modifying other
factors that affect slip \cite{NetoEBBC2005}. Also, excessive roughness may induce turbulence and
instabilities. Furthermore, in some cases increasing surface roughness has little effect on
reducing slip, and may also contribute to sample fracture and structural breakdown
\cite{PerselloCCMP1994, DurairajMSE2009}. Slip can also be controlled by changing the
physicochemical properties of the surface other than roughness, such as wettability, through
chemical treatment or surface coating or by choosing a proper type of material \cite{Schnell1956,
KaoNH1975, ChuraevSS1984, CohenM1985, LukMA1987, HatzikiriakosD1991, Barnes1995, PlucinskiGC1998,
FrancoGB1998, PitHL2000, WallsCSK2003, JosephT2005, SethCB2008, GovardhanSAB2009}.

In the case of highly-viscous fluids and yield-stress materials, adhesives could be used to attach
the sample to the surface to prevent slip \cite{MagninP1987, MagninP1990}. In rheometry, certain
types of rheometric systems, such as the vane geometry, are used to eliminate or reduce slip
\cite{LiddellB1996, FrancoGB1998, Barnes1999, BarnesN2001, RobertsBC2001, Meeten2004, HarteCC2007,
KelessidisM2008}. Another measure to eliminate or reduce wall slip is to increase the dimensions of
the rheometric apparatus, although this could complicate the measurement procedures and may not be
practical in some circumstances due to experimental restrictions \cite{Nguyen1992, Buscall2010}.
Slip may also be controlled by adjusting the flow regime and the surrounding physical conditions,
such as pressure and temperature, and by changing the fluid properties or fluid-solid interaction
through, for instance, adding minute quantities of active substances \cite{ZhuG2001,
GranickZL2003}. On the other side, slip can be enhanced, when it is desirable, by taking opposite
measures such as smoothing the surfaces and employing slip-enhancing additives
\cite{GranickZL2003}.

%SSSSSSSSSSSSSSSSSSSSSSSSSSSSSSSSSSSSS
\section{Conclusions}

Wall slip is a complex phenomenon to which many physical and chemical factors contribute. These
include the type and properties of fluid, flow and surface, as well as the surrounding conditions.
The generally-accepted boundary condition of no-slip at fluid-solid contact should be considered
with caution especially in the non-Newtonian, non-wetting and gaseous systems and at micrometer and
nanometer scales. The acceptance of this condition should be based on thorough assessment in each
individual case.

Serious precautions should be taken, when dealing with flow systems that may suffer from unwanted
slip, to prevent or minimize, or at least detect and correct, this effect to ensure that the
processes are not affected and the measurements are not contaminated with substantial errors.
Yield-stress systems may require special attention as otherwise the detection of yield and the
value of yield-stress, as well as the fluid viscosity and flow rate, could be compromised.

It seems that in many practical situations where wall slip is unwanted, complete eradication of
slip is very hard, if not impossible, to achieve. Therefore, delaying the onset of slip, reducing
its magnitude and neutralizing its effects may be the optimum that the investigator hopes for. The
last resort is to measure the slip velocity and apply corrections to account for the effects of
slip.

Thanks to the complexity of wall slip, the literature of this phenomenon is full of contradicting
views and results, experimental as well as theoretical, and hence many issues will remain
unresolved for long time to come.

%%%%%%%%%%%%%%%%%%%%%%%%%%%%%%%%%%%%%%%%%%%%%%%%%%%%%%%%%%%%%%%%%%%%%%%%%%%%%%%%%%%%%%%%%%%%%%%%
\newpage
\phantomsection \addcontentsline{toc}{section}{References} %
\bibliographystyle{unsrt}
%\bibliography{BiblFluid}

\end{document}